\documentclass[12pt,twoside]{article}

\begin{document}

{\bf What is "system": some decoherence-theory arguments}

\bigskip

\bigskip

{\bf Miroljub Dugi\' c}

\bigskip

\centerline{\it Department of Physics, Faculty of Science,
Kragujevac, Serbia and Montenegro}

\bigskip

theory

methods

\bigskip

{\bf Abstract:} We discuss the possibility of making the {\it
initial} definitions of mutually different (possibly interacting,
or even entangled) systems in the context of decoherence theory.
We point out relativity of the concept of elementary physical
system as well as point out complementarity of the different
possible divisions of a composite system into "subsystems", thus
eventually sharpening the issue of 'what is system'.

\bigskip

{\it 1. Introduction.} Physical system is described by its
parameters (e.g., mass, electric charge etc.) and by the degrees
of freedom properly describing dynamics of the system. In the
classical world, this general scheme seems inevitable.
Nevertheless, in the quantum realm, the things may be different
as we show within the context of the "environment-induced
superselection rules" (or decoherence) theory.

Actually, the task of dividing complex systems into subsystems is
not in general trivial. This fundamental yet a subtle task can be
performed in some generality on the basis of the decoherence
theory, yet bearing certain open questions. E. g.  a composite
system $\mathcal{C}$ may not be divisable in respect to the
arbitrarily defined "degrees of freedom", thus--relative to {\it
these} degrees of freedom--being an {\it elementary} physical
system (likewise the elementary particles). On the other side,
the possible (meaningful) division of $\mathcal{C}$ into
subsystems need {\it not}, in principle, be unique, thus posing
the question of physical reality of the "subsystems" emerging
from the different possible divisions of $\mathcal{C}$. Bearing
in mind that the real systems are usually open systems, the task
of defining "subsystem"  coincides with the task of defining
"system".

The method employed here is elementary yet conceptually
sufficient for addressing the truly fundamental issue of "what is
system" within the context of quantum mechanics of complex
systems.

\bigskip

{\it 2. The problem.} Most of the "quantum paradoxes" start with
the assumption of existence of mutually separable physical
systems. On the other side, quantum holism removes most of the
problems (except on the intuitive level) from the very
beginning--being a consequence of the {\it fully consistent}
quantum mechanical formalism (e.g. of the quantum entanglement).
In the macroscopic domain, however, existence of the
well-defined, mutually separated systems is the very basis of the
physical methods and is actually taken for granted. Thus, in a
sense, transferring the concept of the different systems from the
macroscopic, through the mesoscopic, to the purely
quantum-mechanical domain is at the heart of the problem of the
"transition from quantum to classical" [1-3]. This is also a
problem of practical importance--since, it seems, that in the {\it
realistic} situations, we are able to distinguish between the
different systems (e.g. between the object of measurement and the
measurement instrument).

The possibility of defining mutually independent (compatible)
degrees of freedom (and their conjugate momenta) and of performing
independent ("local") measurements of the observables is a {\it
defining feature} of the different physical systems. The
importance of the issue is rather apparent. E.g. according to
Zurek [2]: "...[quantum mechanical] problems... cannot be even
posed when we refuse to acknowledge the division of the Universe
into separate entities", while "..without the assumption of a
preexisting division of the Universe into individual systems the
requirement that they have a right to their own states cannot be
even formulated". Of course, the rules for defining the preferred
states (e.g. the pointer basis, as well as the pointer
observable) of an open system comes from the foundations of the
decoherence theory. On this basis appeared an early draft [4] of
the problem considered here.

This issue should be distinguished from the  problem of the loss
of individuality of mutually {\it entangled} systems. Actually,
the entangled states refer to the, initially, well-defined
systems: the systems (actually subsystems) are usually assumed
already to be {\it defined}, as well as their state spaces, which
is the basis for defining the entangled states. So, in this
perspective,  the task of answering 'what is system' is a more
fundamental task than investigation of entanglement itself.

Essentially the same problem has recently been addressed e.g. by
Zanardi et al [5] and by Barnum et al [6] (and the references
therein), by considering the different possible operational uses
of entanglement in the quantum information issues. While bearing
some similarity with our results, the results therein presented
are based on the different approaches that is briefly discussed in
Section 5. Here, we employ the foundations of the decoherence
theory and particularly certain recent results in this regard [4,
7, 8].

\bigskip

{\it 3. Separability.} As implicit in the above quotation of
Zurek, the initial definitions of the subsystems {\it make sense
if one can} a posteriori {\it justify these definitions} on the
basis of the {\it occurrence of decoherence}. The relevance of
this statement seems to be apparent in the macroscopic, not
necessarily yet in the mesoscopic or microscopic (fully quantum
mechanical) context.

A detailed analysis of the occurrence of decoherence points out
the condition of {\it separability} (cf. DEF. 1 below) of the
interaction term of the Hamiltonian as the (effective) necessary
condition for the occurrence of decoherence. Investigating the
occurrence of decoherence is truly a subtle task [7-9]. E.g.,
separability of the complete Hamiltonian (of the composite system
"system + environment") is sufficient in this regard [7, 8].
Strong interaction allows the occurrence of decoherence. which
still depends on a number of the details in the model of the
system [7]. On the other side, strong interaction is not necessary
for the occurrence of decoherence [9]. Nevertheless, the
condition of separability in the 'macroscopic context' of the
theory represents an (effective) necessary condition for the
occurrence of decoherence [8].

Now, the {\it separability appears as a condition useful for
defining the 'dividing line'} between the subsystems. Formally,
existence of the subsystems is presented (cf. (1) below) by the
tensor-product symbol, $\otimes$, while  assuming the definitions
of the subsystems through their--implicitly present--degrees of
freedom.

DEF. 1: A bipartite ($1+2$) system's observable $\hat A_{12}$ is
of the {\it separable kind}, if its general form

\begin{equation}
\hat A_{12} = \sum_i \hat B_{1i} \otimes \hat C_{2i},
\end{equation}

\noindent fulfills any of the following, mutually equivalent
conditions: (A) Its spectral form reads $\sum_{i,j} a_{ij} \hat
P_{1i} \otimes \hat \Pi_{2j}$, where appear the (orthogonal)
projectors onto the Hilbert spaces of the two systems; (B) there
exist the two orthonormal bases in the state spaces of the
systems, $\{\vert i \rangle_1\}$, and $\{\vert \alpha
\rangle_2\}$ that diagonalize the observable: $_1\langle i \vert
\hat A_{12} \vert j \rangle_1 = 0, \forall{i \neq j}$, and
$_2\langle \alpha \vert \hat A_{12} \vert \beta \rangle_2 = 0,
\forall{\alpha \neq \beta}$; (C) every pair of the observable of
the system $1$ in (1) mutually commute, and analogously for the
system $2$.

A constructive proof of existence of the general form (1) of a
bipartite system's observable is given in [8]. Depending on the
actual task, any of these definitions of separability may equally
be operationally useful. Needless to say, a bipartite system's
Hamiltonian is subject to DEF. 1.

Thereforre, operationally, investigating separability of the
Hamiltonian gives rise to both [7, 8]: (i) to the superselection
rules defined by the projectors $\{\hat P_{1i}\}$ (when the
system $1$ is considered as the open system), and (ii) to a
definition of the pointer observable and therefore of the
possible pointer basis (or of the "preferred set of states") of
the open system--e.g. the system $1$ in our notation). Having in
mind that the observables, e.g. $\hat B_1$s, are the functions of
the degrees of freedom of the system $1$, the task of
investigating decoherence actually assumes the {\it initially}
well-defined (sub)systems.

Bearing in mind the subtleties concerning the occurrence of
decoherence, we simplify our approach: actually, we employ the
condition of separability as a criterion for making the dividing
line between the subsystems of a composite system.

\bigskip

{\it 4. Quantum relativity of "system".} Usefulness of
separability in the foundations of decoherence theory bears some
subtlety yet. The example of the hydrogen atom is paradigmatic in
the following sense. The composite system "hydrogen atom (HA)" is
originally defined by the Hamiltonian:

\begin{equation}
\hat H = \hat T_e \otimes \hat I_p + \hat I_e \otimes \hat T_p +
\hat V_{Coul},
\end{equation}

\noindent where the Coulomb interaction couples the positions of
the electron (subscript $e$) and the proton (subscript $p$),
bearing obvious notation. Having in mind the definition of
separability (Section 3), it is straightforward to prove
non-separability of $\hat H$ yet separability of the Coulomb
interaction\footnote{The {\it strength} of the separable
interaction term gives rise to the bound states and the
interpretation of  HA as distinguished in the body text.}.

However, the proper {\it canonical transformations} of the degrees
of freedom give another, {\it separable form} of $\hat H$; even
more, each single term is (apparently) of the separable kind:

\begin{equation}
\hat H = \hat T_{CM} \otimes \hat I_R + \hat I_{CM} \otimes \hat
T_R + \hat I_{CM} \otimes V_{Coul} (\hat r_R),
\end{equation}

\noindent where $CM$ stands for the "center of mass" and $R$ for
the "relative particle" system; $r_R \equiv \vert \vec r_e - \vec
r_p \vert$.

In the {\it context} of our considerations, these well-known
transformations give rise to the following observation. The
composite system HA is decomposable\footnote{Here, we do {\it
not} assume the occurrence of decoherence in HA--the proton is
much too small in order to play the role of the environment for
the electron. We just formally employ the separability of the
Coulomb interaction to point out consistency of our
considerations.} into the pair of the quantum particles $(e, p)$,
stemming from separability of the {\it strong} Coulomb
interaction (cf. footnote 1). On the other side, the form (3) of
the Hamiltonian refers to the new, also well known, division of
HA: the system now reads "Center of mass + relative particle"
($CM + R$); certainly, $e+p = \mathcal{C} = CM+R $. Due to the
small mass-ratio--$m_e/m_p \ll 1$--it is (semiclassicly) allowed
to "identify" $CM$ with $p$ and $R$ with $e$. Nevertheless, in
general, this identification is not physically reasonable, as we
show in the sequel. From this example, we learn:

\noindent {\it the choice of the degrees of freedom may redefine
the Hamiltonian separability, thus} (cf. Section 3) {\it directly
referring to the issue of putting the dividing line between the
(sub)systems}.

Let us first briefly consider the case of totally nonseparable
Hamiltonian. That is, we assume that a given Hamiltonian can not
be (re)written in a separable form by the use of any (linear)
canonical transformations. As to the told in Sections 2 and 3,
then one can not define a dividing line between the "subsystems"
of the composite system defined by the Hamiltonian considered.
Then, it seems we are forced to consider the system {\it
undivisable}, thus resembling the concept of elementarity of the
quantum particles. Physically, a definition of the subsystems in
this case is artificial, and the measurements of the "subsystems'
observables" is nothing but the measurements of the observables
of the composite system, not yet interpretable in terms of the
observables of the well-defined subsystems.

As a counterexample, let us analyse the following possibility. A
Hamiltonian is separable {\it relative} to a set of the "degrees
of freedom" (and their conjugate momenta), $(\hat x_{Ai}, \hat
p_{Aj}  ; \hat \xi_{Bm}, \hat \pi_{Bn})$, thus defining a
division of the composite system as $\mathcal{C} = \mathcal{A} +
\mathcal{B}$; by definition, $[\hat x_{Ai}, \hat p_{Aj}] = \imath
\hbar \delta_{ij}$ (and analogously for $\mathcal{B}$), while
$[\hat x_{Ai}, \hat \xi_{Bm}] = 0$ and $[\hat x_{Ai}, \hat
\pi_{Bn}] = 0$ (and analogously for $\hat p_A$s). But, suppose
that the same Hamiltonian can be rewritten in a separable form
relative to another (analogous) set of the "degrees of freedom",
$(\hat X_{Dp}, \hat P_{Dq} ; \hat \zeta_{E\alpha}, \hat
\Pi_{E\beta})$, thus giving rise to another possible division of
the composite system, $\mathcal{C} = \mathcal{D} + \mathcal{E}$.
By the assumption: the two sets of the degrees of freedom are
mutually related by the linear canonical transformations

\begin{equation}
\hat \zeta_{E\alpha} = f_{\alpha} (\hat x_{Ai}, \hat p_{Aj}; \hat
\xi_{Bm}, \hat \pi_{Bn}), \quad \hat \Pi_{E\beta} = g_{\beta}
(\hat x_{Ai}, \hat p_{Aj}; \hat \xi_{Bm}, \hat \pi_{Bn}),
\end{equation}

\noindent and analogously for the subsystem $\mathcal{D}$, while
assuming the inverse is also defined. Needless to say, the
measurements of e.g. $\hat x_{Ai}$, or $\hat \zeta_{Dp}$, may be
interpreted as the measurements of the observables of the
composite system. Yet, the supposed separabilities of the
Hamiltonian allow the interpretations in terms of the subsystems,
again bearing some subtlety.

In general, the measurements of e.g. the observables of
$\mathcal{A}$  partly reveal, yet {\it quantum mechanically
undetermined} values of the observables of both $\mathcal{D}$ and
$\mathcal{E}$--due to (4), one may obtain e.g. $[\hat x_{Ai}, \hat
\zeta_{D\alpha}] \neq 0$. As a consequence: the inverse of (4) can
not be used for determining the definite values of the
observables of $\mathcal{D}$ and $\mathcal{E}$--in
contradistinction with the macroscopic experience. On the other
side, only the measurements of $\mathcal{A}$ and $\mathcal{B}$ (of
$\mathcal{D}$ and $\mathcal{E}$) do not mutually interfere,
referring to the mutually compatible observables. Therefore, the
measurements of the observables of the "subsystems" belonging to
the different divisions do not make sense, while the measurements
referring to the observables of the subsystems belonging to the
{\it same division} of the composite system are physically
reasonable. Needless to say, the later is in agreement with the
standard, general procedure we have learnt in the "classical
domain" (Section 2--cf. also Section 5). As a consequence, the two
possible divisions may refer to the two, mutually complementary,
possible entanglements in the system $\mathcal{C}$: $\sum_i c_i
\vert \psi_i \rangle_A \otimes \vert \chi_i \rangle_B$, and
$\sum_j d_j \vert \psi_j \rangle_D \otimes \vert \phi_j
\rangle_E$ (compare to [5, 6]).

As long as the composite system may be considered to be {\it
isolated}, the two different divisions as described above seem
mutually equivalent for an independent observer. This, however,
need not be the case for an open composite system, as discussed
in [10].

It is probably obvious: a definition of e.g. subsystem
$\mathcal{A}$ makes sense {\it if and only if} the subsystem
$\mathcal{B}$ is {\it simultaneously} defined. This is both a
mathematical consequence of the canonical transformations as well
as physically a reasonable notion.

Therefore, the concept of {\it elementarity} as well as of a {\it
subsystem} are {\it relative}; as to the later, bearing in mind
that the real systems are usually open, the relativity of
"subsystem" actually means relativity of the basic physical
concept of "system".

\bigskip

{\it 5. Discussion.} Depending on the context, we hope the use of
the term "separability" is clear--since it applies to both, the
total Hamiltonian as well as to any (e.g. interaction) term of the
Hamiltonian. Since this is not substantial, we do not explicitly
distinguish between these possibilities throughout the paper. So,
for simplicity, we assume that '(non)separability' (non)defines
the 'dividing line'--which is the main operational tool of our
analysis.

Originally, the problem of "what is system" (Section 2) stems
from the "macroscopic context" of the decoherence theory [1-3]. It
seems that, nowadays, the physicists are ready to accept the
"undivised universe" on the truly microscopic, and partly on the
"mesoscopic" scale. Yet, in the context of the "macroscopic
considerations", it seems unplausible to start with such a
hypothesis [1, 3]. Therefore, our conclusions mainly refer to the
macroscopic context of the decoherence theory; investigating their
extension towards the fully quantum mechanical scales remains an
open task of our analysis. To this end, the above distinguished
use of the condition of separability may be employed as a
(plausible) working hypothesis.

Following the fundamentals of the decoherence theory (yet mainly
in the macroscopic context), we have argued that the condition of
separability of the Hamiltonian may serve as a criterion for
making a "division line" between the susbsystems of a composite
system. As we here emphasize, this reasonable approach gives
automatically rise to the possibility of defining the subsystems,
through a {\it definition} of the degrees of freedom (and their
conjugate momenta) that are {\it based on the condition of
separability} of the Hamiltonian. We have also seen that the
condition of separability is consistent with our macroscopic
experience: e.g. the measurements on $\mathcal{B}$
($\mathcal{A}$) may be performed {\it independently} on the
measurements on the subsystem $\mathcal{A}$ ($\mathcal{B}$); as a
benefit, the subsystem $\mathcal{B}$ ($\mathcal{A}$) may be
defined {\it only simultaneously} with the subsystem $\mathcal{A}$
($\mathcal{B}$). The different divisions of the composite system
may bear quantum mechanical complementarity, being mutually
exclusive divisions of a composite system. This way, the problem
of "what is system" seems to be sharpened, and particularly {\it
reduced} to the following problem: "as to what extent, one may
ascribe the physical reality to the different divisions of a
composite system, especially in the 'macroscopic' context".
Needless to say, much remains yet to be done in this respect. To
this end, the example of the hydrogen atom is useful both, as an
example of applicability of our general models to the {\it
realistic systems} as well as an example exhibiting the
subtleties in the same concern. As to the later, the strength and
separability of the Coulomb interactions relative to the degrees
of freedom of the pair $(e, p)$ justify the division of the atom
into the pair "electron + proton" (cf. footnotes 1 and 2).
Nevertheless, in general, identification of the subsystems
referring to the different divisions of the composite system
(semi-classically plausible identification: $CM = p$, and $R = e$
in HA ) are physically nonjustifiable--complementarity of the
different divisions of a composite system.

Once made, a division of a composite system $\mathcal{C}$ can be
straightforwardly extended (further "coarse graining" of
$\mathcal{C}$) in accordance with the above criteria--e.g.
$\mathcal{A} = \mathcal{A}_1 + \mathcal{A}_2 + \dots$.

Finally, our discussion and conclusions are applicable virtually
to any complex quantum system. However, its relevance to the
realistic systems remains yet to be investigated in the purely
quantum-mechanical ("microscopic") as well as in the mesoscopic
context. To this end, it is interesting to compare our approach
and conclusions with the approach of Zanardi et al [5] and Barnum
et al [6]. A common element of [5] and our considerations is the
notion on the importance of interaction in the composite system in
defining the subsystems. While this is a conclusion in [5], we
still use this plausible notion (stemming from the decoherence
theory) in developing the general models of our analysis. The
approach of Zanardi et al [5] is based on certain axioms
referring to the "experimentally accessible observables", which
is yet an open issue of our approach . Our approach is
characterized by the pointing out separability as an operational
tool in defining the subsystems, yet bearing possibly some
restrictions onto the "macroscopic context". On the other side,
rejecting the "reference to a preferred subsystem decomposition"
of a composite system, Barnum et al [6] seem essentially to point
to the relativity of the concept of "subsystem"--in analogy with
our conclusion. However, being an operational analysis of
entanglement, their paper does not directly tackle the issue of
"what is system". Nevertheless, we believe, that the conclusions
of [5, 6] are consistent with our conclusions, which still follow
from the foundations of the decoherence theory.

\bigskip

The author thanks the Ministry of Science and Environmental
Protection, Serbia, for financial support.

\bigskip

[1] D. Giulini, E. Joos, C. Kiefer, J. Kupsch, I.-O. Stamatescu
and H. D. Zeh, "Decoherence and the Appearance of a Classical
world in Quantum Theory" (Springer, Berlin, 1996)

[2] W. H. Zurek, Prog. Theor. Phys. {\bf 89}, 281 (1993)

[3] W. H. Zurek, Phys. Today {\bf 48}, 36 (1991)

[4] M. Dugi\' c, eprint arXiv quant-ph/9903037

[5] P. Zanardi, D. A. Lidar and S. Lloyd, Phys. Rev. Lett. {\bf
92}, 060402 (2004)

[6] H. Barnum, E. Knill, G. Ortiz, R. Somma and L. Viola, eprint
arxiv quant-ph/ 0305023

[7] M. Dugi\' c, Physica Scripta {\bf 53}, 9 (1996)

[8] M. Dugi\' c, Physica Scripta {\bf 56}, 560 (1997)

[9] J. P. Paz and W. H. Zurek, Phys. Rev. Lett. {\bf 82}, 5181
(1999)

[10] M. Dugi\' c, M. M. \' Cirkovi\' c and D. Rakovi\' c, Open
Systems and Information Dynamics  {\bf 9}, 153 (2002)

\end{document}